%% file: lat98.tex
\documentstyle[twoside,fleqn,espcrc2]{article}

\newcommand{\AmS}{{\protect\the\textfont2
  A\kern-.1667em\lower.5ex\hbox{M}\kern-.125emS}}

\title{Distinguishing J=4 from J=0 on a cubic lattice}

\author{R. Johnson
        and 
        M. Teper \\
\vspace{0.1in}
        Theoretical Physics, University of Oxford, \\
        1 Keble Road, Oxford OX1 3NP,  United Kingdom}
       
\begin{document}


\maketitle

\section{THE PROBLEM}

On a square lattice (we consider here
a D=2+1 lattice for simplicity) the exact rotational symmetry is
confined to rotations that are integer multiples of $\pi/2$. 
If we attempt to build operators of spin J from basic components 
that are identical -- that is, related by a lattice symmetry --  
then there is an ambiguity:
\begin{equation}
    \exp(iJ\theta) \equiv \exp(iJ^{\prime}\theta)  \ \ \ 
\forall \theta = {{n\pi}\over 2} 
\end{equation}
if
\begin{equation}
    J^{\prime} = J + 4 N   \ \ \ \forall N
\end{equation}

Thus, for example, we cannot distinguish J=0 from J=4.

It is conventional to assign the lowest value of continuum J
to such an operator. However even
if this choice is correct for the lightest state of a given `J'
-- see the discussion in \cite{Teper98} --
the ambiguity cannot be avoided once we start calculating 
several excitations of the same `J'.

Of course this is a self-inflicted problem. As $ a\to 0$,
we know that we recover full rotational invariance on physical
length scales. 
Therefore we should be able to construct operators that
become as close as we like to spin J, for any J, as
$a\to 0$.
This will require using linear combinations of operators that
are $approximate$ rotations of each other, where the approximation
is such that it becomes exact as $ a\to 0$.

In this poster we are going to explore how well the simplest 
embodiment of this idea works. We shall do so by attempting to
calculate the J=0 and J=4 glueball masses in the 2+1 
dimensional SU(2) lattice gauge theory.

\section{FIRST STEPS ...}

We start with some closed loop on the lattice that is
symmetric about the x-axis. Call it $\phi_{Ax}$. There
is a corresponding loop about the y-axis, $\phi_{Ay}$,
which is identical in the sense that e.g.
\begin{equation}
 \langle  \phi_{Ay}(t) \phi_{Ay}(0) \rangle
=
 \langle  \phi_{Ax}(t) \phi_{Ax}(0) \rangle
\ \ \ \ \forall t
\end{equation}
If we sum these two, $\phi_A =\phi_{Ax}+\phi_{Ay}$, then 
we obtain an operator  $\phi_A$ that we would normally 
call `J=0'. (As usual we always take zero-momentum sums
of such operators.)

We now construct a loop that is symmetric around the diagonal
which is at $\pi/4$ to the x-axis. We choose a loop that
`looks' as though it is roughly a rotation of  $\phi_{Ax}$.
There will be an identical loop that is rotated by $\pi/2$.
Summing these two loops gives us the diagonal operator $\phi_D$.

Clearly our trial J=0 operator is going to be
\begin{equation}
\phi_{J=0} = \phi_{A} + c\phi_{D}
\end{equation}
and our trial J=4 operator is going to be
\begin{equation}
\phi_{J=4} = \phi_{A} - c\phi_{D}
\end{equation}
(since exp(iJ$\theta$)=-1 for J=4 and $\theta=\pi/4$.)

If $\phi_A$ and $\phi_D$ were indeed identical (up
to a rotation) then we would take c=1 and then our `J=0'
state would contain no J=4 and vice-versa. However they are not
identical
and so we will choose c so as to enhance whatever approximate
identity they possess e.g.
\begin{equation}
c^2 \langle  \phi_D(t) \phi_D(0) \rangle
\simeq
 \langle  \phi_A(t) \phi_A(0) \rangle
\ \ \ \ \forall t
\end{equation}

We now give the simplest example of such operators. Let $U_j$ be the 
matrix on the link in direction $j$. I suppress the site index. Then 
our example consists of the following path ordered products
\begin{equation}
\phi_A = 
Tr\{U_xU_xU_yU^{\dagger}_xU^{\dagger}_xU^{\dagger}_y\}
\end{equation}
and
\begin{equation}
\phi_D =
Tr\{U_xU_yU_xU_y
U^{\dagger}_xU^{\dagger}_yU^{\dagger}_xU^{\dagger}_y
\} 
\end{equation}
with, in each case, the addition of the operator that is rotated by
an angle of $\pi/2$. So the basic on-axis operator is a $1\times 2$
rectangular loop, and the corresponding `diagonal' operator is
an ordered product of 2 plaquettes. We can obviously extend
this to $1\times n$ rectangular loops and to diagonal  operators
that are ordered products of $n$ plaquettes. 

This operator is only useful if  $\phi_D$ is close to
being identical to $\phi_A$ (up to a normalisation) in
which case the operators are close to being $\pi/4$ rotations
of each other and we can form good J=0 and J=4 combinations.
In Fig.1 we compare these operators, using `twice-blocked'
link matrices
\cite{block}.
(The normalisation $c$ in eqns(4,5) has been chosen so that the
vacuum-subtracted correlation functions at $t=0$ are
equal.) We observe that the
on-axis and diagonal operators are indeed `close' to
being identical and so provide at least some kind of
starting point for distinguishing $J=0$ from $J=4$.

\begin	{figure}[t]
\begin	{flushright}
\leavevmode
\input	{plot_corr12}
\end	{flushright}
\caption{On-axis ($\bullet$) and diagonal ($\circ$)
$1\times 2$ loop vacuum-subtracted correlations.}
\end 	{figure}
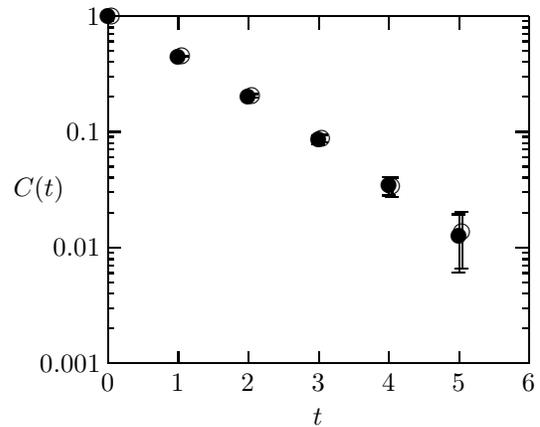

\section{A CALCULATION}

As a starting basis of operators we 
use $1 \times 2$, $1 \times 3$,  $1 \times 4$
and   $2 \times 4$ loops, with the diagonal loops
chosen in the obvious way. We construct these loops
not just out of the elementary lattice links but
also out of blocked links (to enhance the overlaps
onto the lightest states). In practice we shall use once, 
twice and thrice blocked links so that our initial
basis contains $4 \times 3=12$ operators.
As seen above, choosing the constant c in eqns(4,5)
so that
\begin{equation}
 c^2 \langle  \Phi_D(t) \Phi_D(0) \rangle
=
 \langle  \Phi_A(t) \Phi_A(0) \rangle
\end{equation}
for the vacuum-subtracted operators, with t=0 
(or indeed t=a), works quite well.
We will, from now on, choose to use t=0,
and all the operators
we use shall be vacuum subtracted and normalised
so the $t=0$ correlators are unity.

The exploratory calculations shown here are
obtained on a modest run of 5000 sweeps on a $24^3$ lattice at 
$\beta=9$ in the D=2+1 SU(2) lattice gauge theory.

If our operators really were J=0 and J=4
-- which can only occur in the continuum limit --
then they would be mutually orthogonal:
\begin{equation}
\langle\Phi_{J=0}(0) \Phi_{J=4}(0)\rangle = 0. 
\end{equation}
We calculate these overlaps and find that the worst operator
(in the sense of having the largest such overlap)
is the one based on the $1\times 2$ loop. Hence
we discard it from now on, so that our basis is 
reduced to 9 operators. For these operators we
find that the overlaps in eqn(10) are very small compared
to the overlaps
$\langle\Phi_{J=0} \Phi^{\prime}_{J=0}\rangle$. This
confirms that our procedure is working reasonably well.

How do we extract a mass? We know from positivity that  
\begin{equation}
\langle  \Phi_J(t) \Phi_J(0) \rangle
\leq e^{-m_J t}
\end{equation}
where $m_J$ is the lightest mass (or energy).

So given a set of operators $\Phi^i_J$; i=1,..,N,
a variational estimate of the mass is
provided by 
\begin{equation}
m_J = \min_{i} m^i_{J,eff}(t)
\end{equation}
where we have defined a set of effective masses by
\begin{equation}
m^i_{J,eff}(t) = - {1\over t} \ln\{
\langle  \Phi^i_J(t) \Phi^i_J(0) \rangle\}.
\end{equation}

In Fig.2 we show the values of
$m^i_{eff}(t=a)$ for our nine J=0 and J=4
operators. We use t=a both because the
effective masses become rapidly less accurate
if we increase t; and also because we know
that at large enough t the lightest scalar mass will
dominate. That is to say, a good `J=4' operator will have 
two quite distinct effective mass plateaux: the one at
lower t will give $m_{J=4}$ while the one at
larger t will give  $m_{J=0}$ (if it is lighter).
As a$\to 0$ the first plateau will extend to
ever larger values of t. In this preliminary test of
the method we do not attempt to illustrate such behaviour.  

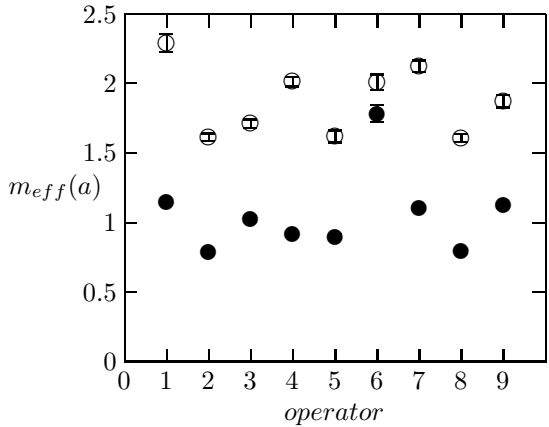
\begin	{figure}[t]
\begin	{flushright}
\leavevmode
\input	{plot_meff}
\end	{flushright}
\caption{$t=a$ effective masses from $J=0$
($\bullet$) and $J=4$ ($\circ$) correlators.}
\end 	{figure}

In our plot we see a distinct seperation between
the J=0 and J=4 effective masses at t=a and, applying
the above variational criterion, we obtain the
following mass estimates (in lattice units):
\begin{equation}
m(J=0) = 0.786(12)
\end{equation}
\begin{equation}
m(J=4) = 1.607(27)
\end{equation}
The $J=0$ mass is close to the expected value
\cite{Teper98}.
Is the lightest $J=4$ state a glueball or a state
composed of 2 $J=0$ glueballs in an $L=4$ partial wave?
The latter is unlikely: the $L=4$ suppression
near the branch point of the cut should lead to an effective
mass that is considerably above the energy at the
branch point $E=2m(J=0)$. In addition we expect our 
single trace operators to have a small projection
on states composed of two colour singlets. Evidence for 
these expectations comes from earlier 
numerical work
\cite{cuts}.
Although all this will eventually need explicit investigation,
in this preliminary study we assume it to be so and that
what we have calculated is the lightest $J=4$ glueball mass.
 
A better variational calculation would be to work with
all linear combinations of the N(=9) $\Phi^i_J$.
This is discussed in
\cite{future}.

Of course, simple as all this is in principle, in
practice it rapidly becomes something of a Heath
Robinson construction if one wants to extend it
to arbitrary J. 
What one wants is a sytematic, general procedure 
that does not require tailoring for the individual
`bulges' of each J. For example one can proceed as follows.      
As $a\to 0$ the density of lattice sites within two
concentric circles of radii $r$ and $r^\prime$, where
these radii are fixed in physical units, becomes arbitrarily dense.
This will even be the case if we choose $r \to r^\prime$ in
physical units, as long as $|r-r^\prime|/a \to \infty$.
In that case we can construct triangular operators 
(constructed out of some well-chosen gluonic propagators
connecting the centre to 2 sites between $r$ and $r^\prime$) 
that become rotations of each other for arbitrary angles of rotation
as $a \to 0$.
These can then be used as part of a systematic procedure for
constructing operators of arbitrary spin J as $a\to 0$.
This approach is described in
\cite{future}.

\end{document}

%% file: plot_corr12.tex
\setlength{\unitlength}{0.240900pt}
\ifx\plotpoint\undefined\newsavebox{\plotpoint}\fi
\sbox{\plotpoint}{\rule[-0.200pt]{0.400pt}{0.400pt}}%
\begin{picture}(825,584)(0,0)
\font\gnuplot=cmr10 at 12pt
\gnuplot
\sbox{\plotpoint}{\rule[-0.200pt]{0.400pt}{0.400pt}}%
\put(120.0,31.0){\rule[-0.200pt]{4.818pt}{0.400pt}}
\put(108,31){\makebox(0,0)[r]{{$0.001$}}}
\put(761.0,31.0){\rule[-0.200pt]{4.818pt}{0.400pt}}
\put(120.0,86.0){\rule[-0.200pt]{2.409pt}{0.400pt}}
\put(771.0,86.0){\rule[-0.200pt]{2.409pt}{0.400pt}}
\put(120.0,118.0){\rule[-0.200pt]{2.409pt}{0.400pt}}
\put(771.0,118.0){\rule[-0.200pt]{2.409pt}{0.400pt}}
\put(120.0,141.0){\rule[-0.200pt]{2.409pt}{0.400pt}}
\put(771.0,141.0){\rule[-0.200pt]{2.409pt}{0.400pt}}
\put(120.0,158.0){\rule[-0.200pt]{2.409pt}{0.400pt}}
\put(771.0,158.0){\rule[-0.200pt]{2.409pt}{0.400pt}}
\put(120.0,173.0){\rule[-0.200pt]{2.409pt}{0.400pt}}
\put(771.0,173.0){\rule[-0.200pt]{2.409pt}{0.400pt}}
\put(120.0,185.0){\rule[-0.200pt]{2.409pt}{0.400pt}}
\put(771.0,185.0){\rule[-0.200pt]{2.409pt}{0.400pt}}
\put(120.0,195.0){\rule[-0.200pt]{2.409pt}{0.400pt}}
\put(771.0,195.0){\rule[-0.200pt]{2.409pt}{0.400pt}}
\put(120.0,205.0){\rule[-0.200pt]{2.409pt}{0.400pt}}
\put(771.0,205.0){\rule[-0.200pt]{2.409pt}{0.400pt}}
\put(120.0,213.0){\rule[-0.200pt]{4.818pt}{0.400pt}}
\put(108,213){\makebox(0,0)[r]{{$0.01$}}}
\put(761.0,213.0){\rule[-0.200pt]{4.818pt}{0.400pt}}
\put(120.0,268.0){\rule[-0.200pt]{2.409pt}{0.400pt}}
\put(771.0,268.0){\rule[-0.200pt]{2.409pt}{0.400pt}}
\put(120.0,300.0){\rule[-0.200pt]{2.409pt}{0.400pt}}
\put(771.0,300.0){\rule[-0.200pt]{2.409pt}{0.400pt}}
\put(120.0,323.0){\rule[-0.200pt]{2.409pt}{0.400pt}}
\put(771.0,323.0){\rule[-0.200pt]{2.409pt}{0.400pt}}
\put(120.0,340.0){\rule[-0.200pt]{2.409pt}{0.400pt}}
\put(771.0,340.0){\rule[-0.200pt]{2.409pt}{0.400pt}}
\put(120.0,355.0){\rule[-0.200pt]{2.409pt}{0.400pt}}
\put(771.0,355.0){\rule[-0.200pt]{2.409pt}{0.400pt}}
\put(120.0,367.0){\rule[-0.200pt]{2.409pt}{0.400pt}}
\put(771.0,367.0){\rule[-0.200pt]{2.409pt}{0.400pt}}
\put(120.0,377.0){\rule[-0.200pt]{2.409pt}{0.400pt}}
\put(771.0,377.0){\rule[-0.200pt]{2.409pt}{0.400pt}}
\put(120.0,387.0){\rule[-0.200pt]{2.409pt}{0.400pt}}
\put(771.0,387.0){\rule[-0.200pt]{2.409pt}{0.400pt}}
\put(120.0,395.0){\rule[-0.200pt]{4.818pt}{0.400pt}}
\put(108,395){\makebox(0,0)[r]{{$0.1$}}}
\put(761.0,395.0){\rule[-0.200pt]{4.818pt}{0.400pt}}
\put(120.0,450.0){\rule[-0.200pt]{2.409pt}{0.400pt}}
\put(771.0,450.0){\rule[-0.200pt]{2.409pt}{0.400pt}}
\put(120.0,482.0){\rule[-0.200pt]{2.409pt}{0.400pt}}
\put(771.0,482.0){\rule[-0.200pt]{2.409pt}{0.400pt}}
\put(120.0,505.0){\rule[-0.200pt]{2.409pt}{0.400pt}}
\put(771.0,505.0){\rule[-0.200pt]{2.409pt}{0.400pt}}
\put(120.0,522.0){\rule[-0.200pt]{2.409pt}{0.400pt}}
\put(771.0,522.0){\rule[-0.200pt]{2.409pt}{0.400pt}}
\put(120.0,537.0){\rule[-0.200pt]{2.409pt}{0.400pt}}
\put(771.0,537.0){\rule[-0.200pt]{2.409pt}{0.400pt}}
\put(120.0,549.0){\rule[-0.200pt]{2.409pt}{0.400pt}}
\put(771.0,549.0){\rule[-0.200pt]{2.409pt}{0.400pt}}
\put(120.0,559.0){\rule[-0.200pt]{2.409pt}{0.400pt}}
\put(771.0,559.0){\rule[-0.200pt]{2.409pt}{0.400pt}}
\put(120.0,569.0){\rule[-0.200pt]{2.409pt}{0.400pt}}
\put(771.0,569.0){\rule[-0.200pt]{2.409pt}{0.400pt}}
\put(120.0,577.0){\rule[-0.200pt]{4.818pt}{0.400pt}}
\put(108,577){\makebox(0,0)[r]{{$1$}}}
\put(761.0,577.0){\rule[-0.200pt]{4.818pt}{0.400pt}}
\put(120.0,31.0){\rule[-0.200pt]{0.400pt}{4.818pt}}
\put(120,19){\makebox(0,0){\shortstack{\\ \\ \\ {$0$}}}}
\put(120.0,557.0){\rule[-0.200pt]{0.400pt}{4.818pt}}
\put(230.0,31.0){\rule[-0.200pt]{0.400pt}{4.818pt}}
\put(230,19){\makebox(0,0){\shortstack{\\ \\ \\ {$1$}}}}
\put(230.0,557.0){\rule[-0.200pt]{0.400pt}{4.818pt}}
\put(340.0,31.0){\rule[-0.200pt]{0.400pt}{4.818pt}}
\put(340,19){\makebox(0,0){\shortstack{\\ \\ \\ {$2$}}}}
\put(340.0,557.0){\rule[-0.200pt]{0.400pt}{4.818pt}}
\put(451.0,31.0){\rule[-0.200pt]{0.400pt}{4.818pt}}
\put(451,19){\makebox(0,0){\shortstack{\\ \\ \\ {$3$}}}}
\put(451.0,557.0){\rule[-0.200pt]{0.400pt}{4.818pt}}
\put(561.0,31.0){\rule[-0.200pt]{0.400pt}{4.818pt}}
\put(561,19){\makebox(0,0){\shortstack{\\ \\ \\ {$4$}}}}
\put(561.0,557.0){\rule[-0.200pt]{0.400pt}{4.818pt}}
\put(671.0,31.0){\rule[-0.200pt]{0.400pt}{4.818pt}}
\put(671,19){\makebox(0,0){\shortstack{\\ \\ \\ {$5$}}}}
\put(671.0,557.0){\rule[-0.200pt]{0.400pt}{4.818pt}}
\put(781.0,31.0){\rule[-0.200pt]{0.400pt}{4.818pt}}
\put(781,19){\makebox(0,0){\shortstack{\\ \\ \\ {$6$}}}}
\put(781.0,557.0){\rule[-0.200pt]{0.400pt}{4.818pt}}
\put(120.0,31.0){\rule[-0.200pt]{159.235pt}{0.400pt}}
\put(781.0,31.0){\rule[-0.200pt]{0.400pt}{131.531pt}}
\put(120.0,577.0){\rule[-0.200pt]{159.235pt}{0.400pt}}
\put(12,304){\makebox(0,0){{{$C(t)$}}}}
\put(450,-53){\makebox(0,0){{{$t$}}}}
\put(120.0,31.0){\rule[-0.200pt]{0.400pt}{131.531pt}}
\put(120,577){\circle*{24}}
\put(230,513){\circle*{24}}
\put(340,450){\circle*{24}}
\put(451,383){\circle*{24}}
\put(561,311){\circle*{24}}
\put(671,232){\circle*{24}}
\put(120,577){\usebox{\plotpoint}}
\put(110.0,577.0){\rule[-0.200pt]{4.818pt}{0.400pt}}
\put(110.0,577.0){\rule[-0.200pt]{4.818pt}{0.400pt}}
\put(230.0,512.0){\rule[-0.200pt]{0.400pt}{0.482pt}}
\put(220.0,512.0){\rule[-0.200pt]{4.818pt}{0.400pt}}
\put(220.0,514.0){\rule[-0.200pt]{4.818pt}{0.400pt}}
\put(340.0,448.0){\rule[-0.200pt]{0.400pt}{0.964pt}}
\put(330.0,448.0){\rule[-0.200pt]{4.818pt}{0.400pt}}
\put(330.0,452.0){\rule[-0.200pt]{4.818pt}{0.400pt}}
\put(451.0,376.0){\rule[-0.200pt]{0.400pt}{3.132pt}}
\put(441.0,376.0){\rule[-0.200pt]{4.818pt}{0.400pt}}
\put(441.0,389.0){\rule[-0.200pt]{4.818pt}{0.400pt}}
\put(561.0,295.0){\rule[-0.200pt]{0.400pt}{6.986pt}}
\put(551.0,295.0){\rule[-0.200pt]{4.818pt}{0.400pt}}
\put(551.0,324.0){\rule[-0.200pt]{4.818pt}{0.400pt}}
\put(671.0,174.0){\rule[-0.200pt]{0.400pt}{21.922pt}}
\put(661.0,174.0){\rule[-0.200pt]{4.818pt}{0.400pt}}
\put(661.0,265.0){\rule[-0.200pt]{4.818pt}{0.400pt}}
\put(126,577){\circle{24}}
\put(236,514){\circle{24}}
\put(346,452){\circle{24}}
\put(456,384){\circle{24}}
\put(566,309){\circle{24}}
\put(676,237){\circle{24}}
\put(126,577){\usebox{\plotpoint}}
\put(116.0,577.0){\rule[-0.200pt]{4.818pt}{0.400pt}}
\put(116.0,577.0){\rule[-0.200pt]{4.818pt}{0.400pt}}
\put(236.0,513.0){\usebox{\plotpoint}}
\put(226.0,513.0){\rule[-0.200pt]{4.818pt}{0.400pt}}
\put(226.0,514.0){\rule[-0.200pt]{4.818pt}{0.400pt}}
\put(346.0,449.0){\rule[-0.200pt]{0.400pt}{1.204pt}}
\put(336.0,449.0){\rule[-0.200pt]{4.818pt}{0.400pt}}
\put(336.0,454.0){\rule[-0.200pt]{4.818pt}{0.400pt}}
\put(456.0,378.0){\rule[-0.200pt]{0.400pt}{2.891pt}}
\put(446.0,378.0){\rule[-0.200pt]{4.818pt}{0.400pt}}
\put(446.0,390.0){\rule[-0.200pt]{4.818pt}{0.400pt}}
\put(566.0,293.0){\rule[-0.200pt]{0.400pt}{7.227pt}}
\put(556.0,293.0){\rule[-0.200pt]{4.818pt}{0.400pt}}
\put(556.0,323.0){\rule[-0.200pt]{4.818pt}{0.400pt}}
\put(676.0,180.0){\rule[-0.200pt]{0.400pt}{21.440pt}}
\put(666.0,180.0){\rule[-0.200pt]{4.818pt}{0.400pt}}
\put(666.0,269.0){\rule[-0.200pt]{4.818pt}{0.400pt}}
\end{picture}

%% file: plot_meff.tex
\setlength{\unitlength}{0.240900pt}
\ifx\plotpoint\undefined\newsavebox{\plotpoint}\fi
\sbox{\plotpoint}{\rule[-0.200pt]{0.400pt}{0.400pt}}%
\begin{picture}(825,584)(0,0)
\font\gnuplot=cmr10 at 12pt
\gnuplot
\sbox{\plotpoint}{\rule[-0.200pt]{0.400pt}{0.400pt}}%
\put(120.0,31.0){\rule[-0.200pt]{4.818pt}{0.400pt}}
\put(108,31){\makebox(0,0)[r]{{$0$}}}
\put(761.0,31.0){\rule[-0.200pt]{4.818pt}{0.400pt}}
\put(120.0,140.0){\rule[-0.200pt]{4.818pt}{0.400pt}}
\put(108,140){\makebox(0,0)[r]{{$0.5$}}}
\put(761.0,140.0){\rule[-0.200pt]{4.818pt}{0.400pt}}
\put(120.0,249.0){\rule[-0.200pt]{4.818pt}{0.400pt}}
\put(108,249){\makebox(0,0)[r]{{$1$}}}
\put(761.0,249.0){\rule[-0.200pt]{4.818pt}{0.400pt}}
\put(120.0,359.0){\rule[-0.200pt]{4.818pt}{0.400pt}}
\put(108,359){\makebox(0,0)[r]{{$1.5$}}}
\put(761.0,359.0){\rule[-0.200pt]{4.818pt}{0.400pt}}
\put(120.0,468.0){\rule[-0.200pt]{4.818pt}{0.400pt}}
\put(108,468){\makebox(0,0)[r]{{$2$}}}
\put(761.0,468.0){\rule[-0.200pt]{4.818pt}{0.400pt}}
\put(120.0,577.0){\rule[-0.200pt]{4.818pt}{0.400pt}}
\put(108,577){\makebox(0,0)[r]{{$2.5$}}}
\put(761.0,577.0){\rule[-0.200pt]{4.818pt}{0.400pt}}
\put(120.0,31.0){\rule[-0.200pt]{0.400pt}{4.818pt}}
\put(120,19){\makebox(0,0){\shortstack{\\ \\ \\ {$0$}}}}
\put(120.0,557.0){\rule[-0.200pt]{0.400pt}{4.818pt}}
\put(186.0,31.0){\rule[-0.200pt]{0.400pt}{4.818pt}}
\put(186,19){\makebox(0,0){\shortstack{\\ \\ \\ {$1$}}}}
\put(186.0,557.0){\rule[-0.200pt]{0.400pt}{4.818pt}}
\put(252.0,31.0){\rule[-0.200pt]{0.400pt}{4.818pt}}
\put(252,19){\makebox(0,0){\shortstack{\\ \\ \\ {$2$}}}}
\put(252.0,557.0){\rule[-0.200pt]{0.400pt}{4.818pt}}
\put(318.0,31.0){\rule[-0.200pt]{0.400pt}{4.818pt}}
\put(318,19){\makebox(0,0){\shortstack{\\ \\ \\ {$3$}}}}
\put(318.0,557.0){\rule[-0.200pt]{0.400pt}{4.818pt}}
\put(384.0,31.0){\rule[-0.200pt]{0.400pt}{4.818pt}}
\put(384,19){\makebox(0,0){\shortstack{\\ \\ \\ {$4$}}}}
\put(384.0,557.0){\rule[-0.200pt]{0.400pt}{4.818pt}}
\put(451.0,31.0){\rule[-0.200pt]{0.400pt}{4.818pt}}
\put(451,19){\makebox(0,0){\shortstack{\\ \\ \\ {$5$}}}}
\put(451.0,557.0){\rule[-0.200pt]{0.400pt}{4.818pt}}
\put(517.0,31.0){\rule[-0.200pt]{0.400pt}{4.818pt}}
\put(517,19){\makebox(0,0){\shortstack{\\ \\ \\ {$6$}}}}
\put(517.0,557.0){\rule[-0.200pt]{0.400pt}{4.818pt}}
\put(583.0,31.0){\rule[-0.200pt]{0.400pt}{4.818pt}}
\put(583,19){\makebox(0,0){\shortstack{\\ \\ \\ {$7$}}}}
\put(583.0,557.0){\rule[-0.200pt]{0.400pt}{4.818pt}}
\put(649.0,31.0){\rule[-0.200pt]{0.400pt}{4.818pt}}
\put(649,19){\makebox(0,0){\shortstack{\\ \\ \\ {$8$}}}}
\put(649.0,557.0){\rule[-0.200pt]{0.400pt}{4.818pt}}
\put(715.0,31.0){\rule[-0.200pt]{0.400pt}{4.818pt}}
\put(715,19){\makebox(0,0){\shortstack{\\ \\ \\ {$9$}}}}
\put(715.0,557.0){\rule[-0.200pt]{0.400pt}{4.818pt}}
\put(120.0,31.0){\rule[-0.200pt]{159.235pt}{0.400pt}}
\put(781.0,31.0){\rule[-0.200pt]{0.400pt}{131.531pt}}
\put(120.0,577.0){\rule[-0.200pt]{159.235pt}{0.400pt}}
\put(12,304){\makebox(0,0){{{$m_{eff}(a)$}}}}
\put(450,-53){\makebox(0,0){{{$operator$}}}}
\put(120.0,31.0){\rule[-0.200pt]{0.400pt}{131.531pt}}
\put(186,281){\circle*{24}}
\put(252,203){\circle*{24}}
\put(318,254){\circle*{24}}
\put(384,231){\circle*{24}}
\put(451,227){\circle*{24}}
\put(517,420){\circle*{24}}
\put(583,272){\circle*{24}}
\put(649,205){\circle*{24}}
\put(715,276){\circle*{24}}
\put(186.0,278.0){\rule[-0.200pt]{0.400pt}{1.445pt}}
\put(176.0,278.0){\rule[-0.200pt]{4.818pt}{0.400pt}}
\put(176.0,284.0){\rule[-0.200pt]{4.818pt}{0.400pt}}
\put(252.0,200.0){\rule[-0.200pt]{0.400pt}{1.204pt}}
\put(242.0,200.0){\rule[-0.200pt]{4.818pt}{0.400pt}}
\put(242.0,205.0){\rule[-0.200pt]{4.818pt}{0.400pt}}
\put(318.0,250.0){\rule[-0.200pt]{0.400pt}{2.168pt}}
\put(308.0,250.0){\rule[-0.200pt]{4.818pt}{0.400pt}}
\put(308.0,259.0){\rule[-0.200pt]{4.818pt}{0.400pt}}
\put(384.0,228.0){\rule[-0.200pt]{0.400pt}{1.204pt}}
\put(374.0,228.0){\rule[-0.200pt]{4.818pt}{0.400pt}}
\put(374.0,233.0){\rule[-0.200pt]{4.818pt}{0.400pt}}
\put(451.0,223.0){\rule[-0.200pt]{0.400pt}{1.927pt}}
\put(441.0,223.0){\rule[-0.200pt]{4.818pt}{0.400pt}}
\put(441.0,231.0){\rule[-0.200pt]{4.818pt}{0.400pt}}
\put(517.0,407.0){\rule[-0.200pt]{0.400pt}{6.504pt}}
\put(507.0,407.0){\rule[-0.200pt]{4.818pt}{0.400pt}}
\put(507.0,434.0){\rule[-0.200pt]{4.818pt}{0.400pt}}
\put(583.0,269.0){\rule[-0.200pt]{0.400pt}{1.445pt}}
\put(573.0,269.0){\rule[-0.200pt]{4.818pt}{0.400pt}}
\put(573.0,275.0){\rule[-0.200pt]{4.818pt}{0.400pt}}
\put(649.0,202.0){\rule[-0.200pt]{0.400pt}{1.445pt}}
\put(639.0,202.0){\rule[-0.200pt]{4.818pt}{0.400pt}}
\put(639.0,208.0){\rule[-0.200pt]{4.818pt}{0.400pt}}
\put(715.0,272.0){\rule[-0.200pt]{0.400pt}{1.927pt}}
\put(705.0,272.0){\rule[-0.200pt]{4.818pt}{0.400pt}}
\put(705.0,280.0){\rule[-0.200pt]{4.818pt}{0.400pt}}
\put(186,532){\circle{24}}
\put(252,383){\circle{24}}
\put(318,405){\circle{24}}
\put(384,471){\circle{24}}
\put(451,385){\circle{24}}
\put(517,470){\circle{24}}
\put(583,495){\circle{24}}
\put(649,382){\circle{24}}
\put(715,440){\circle{24}}
\put(186.0,517.0){\rule[-0.200pt]{0.400pt}{6.986pt}}
\put(176.0,517.0){\rule[-0.200pt]{4.818pt}{0.400pt}}
\put(176.0,546.0){\rule[-0.200pt]{4.818pt}{0.400pt}}
\put(252.0,377.0){\rule[-0.200pt]{0.400pt}{2.891pt}}
\put(242.0,377.0){\rule[-0.200pt]{4.818pt}{0.400pt}}
\put(242.0,389.0){\rule[-0.200pt]{4.818pt}{0.400pt}}
\put(318.0,398.0){\rule[-0.200pt]{0.400pt}{3.132pt}}
\put(308.0,398.0){\rule[-0.200pt]{4.818pt}{0.400pt}}
\put(308.0,411.0){\rule[-0.200pt]{4.818pt}{0.400pt}}
\put(384.0,463.0){\rule[-0.200pt]{0.400pt}{3.613pt}}
\put(374.0,463.0){\rule[-0.200pt]{4.818pt}{0.400pt}}
\put(374.0,478.0){\rule[-0.200pt]{4.818pt}{0.400pt}}
\put(451.0,375.0){\rule[-0.200pt]{0.400pt}{4.577pt}}
\put(441.0,375.0){\rule[-0.200pt]{4.818pt}{0.400pt}}
\put(441.0,394.0){\rule[-0.200pt]{4.818pt}{0.400pt}}
\put(517.0,457.0){\rule[-0.200pt]{0.400pt}{6.022pt}}
\put(507.0,457.0){\rule[-0.200pt]{4.818pt}{0.400pt}}
\put(507.0,482.0){\rule[-0.200pt]{4.818pt}{0.400pt}}
\put(583.0,486.0){\rule[-0.200pt]{0.400pt}{4.336pt}}
\put(573.0,486.0){\rule[-0.200pt]{4.818pt}{0.400pt}}
\put(573.0,504.0){\rule[-0.200pt]{4.818pt}{0.400pt}}
\put(649.0,376.0){\rule[-0.200pt]{0.400pt}{2.891pt}}
\put(639.0,376.0){\rule[-0.200pt]{4.818pt}{0.400pt}}
\put(639.0,388.0){\rule[-0.200pt]{4.818pt}{0.400pt}}
\put(715.0,430.0){\rule[-0.200pt]{0.400pt}{4.818pt}}
\put(705.0,430.0){\rule[-0.200pt]{4.818pt}{0.400pt}}
\put(705.0,450.0){\rule[-0.200pt]{4.818pt}{0.400pt}}
\end{picture}